\documentclass{aa}
\usepackage{graphicx}
\usepackage{txfonts}

\begin{document}
%=======================================================================
% The header of the document:
%=======================================================================
        \title{Rosseland and Planck Mean Opacities for Protoplanetary
               Discs}

        \author{D. Semenov\inst{1}, Th. Henning\inst{2}, M.
                Ch. Helling\inst{4}, Ilgner\inst{3}, 
                \and
                E. Sedlmayr\inst{4}
                }

        \offprints{D. Semenov, dima@astro.uni-jena.de}

        \institute{Astrophysical Institute and University Observatory,
                   Schillerg\"a{\ss}chen 2-3, 07745 Jena, Germany\\
                \and
                   Max-Planck-Institute for Astronomy,
                   K\"onigstuhl 17, 69117 Heidelberg, Germany\\
                \and
                   Institute of Astronomy and Astrophysics, Auf der Morgenstelle
                   10, 72076 T\"{u}bingen, Germany\\
                \and
                   Center for Astronomy and Astrophysics, TU Berlin,
                   Hardenbergstra{\ss}e 36, 10623 Berlin, Germany\\
                  }

        \date{Received <date> / Accepted <date>}
%======================================================================
% The abstract:
%======================================================================
\abstract{In this paper, we present mean gas and dust opacities
relevant to the physical conditions typical of protoplanetary
discs. As the principal absorber for temperatures below $\sim
1\,500$~K, we consider spherical and aggregate dust particles of
various sizes, chemical structure, and porosity, consisting of ice,
organics, troilite, silicates, and iron. For higher temperatures,
ions, atoms, molecules, and electrons are included as the main
opacity sources. Rosseland and
Planck mean opacities are calculated for temperatures between $5$\,K and
$10\,000$\,K and gas densities ranging from $10^{-18}$~g\,cm$^{-3}$ to
$10^{-7}$~g\,cm$^{-3}$. The dependence on the adopted model of
dust grains is investigated. We compare our results with recent
opacity tables and show how different opacity models affect
the calculated hydrodynamical structure of accretion discs.
        }
%======================================================================
% Print created title:
%======================================================================
        \authorrunning{D.~Semenov et al.}
        \titlerunning{Opacities for protoplanetary discs}
        \maketitle
%======================================================================
% I. Introduction:
%======================================================================
\section{Introduction}
Recently, significant progress toward the understanding of the
possible composition and properties of dust grains and gas species in
many astrophysical environments has been achieved. For instance,
(sub)millimetre observations of molecular lines provided basic
information about chemical composition and dynamical properties of the
gas in discs around pre-main-sequence stars and young stellar objects
(e.g., Bujarrabal et al.~\cite{Bea97}; Olofsson, Liseau, \&
Brandeker~\cite{OLB01}; Thi et al.~\cite{Tea01}; Pi{\'e}tu et al.~\cite{PDK03}). 
The infrared-to-millimetre continuum observations
of such environments constrain the properties of dust grains and can be
used to estimate masses and thermal structure of the objects (e.g., Bouwman et
al.~\cite{Bea00}; Boogert, Hogerheijde, \& Blake~\cite{BHB02}; Tuthill et 
al.~\cite{Tea02}). Finally, experimental
studies on the formation and spectra of various gaseous species (e.g.,
Butler et al.~\cite{BLP01}; Sanz, McCarthy, \& Thaddeus~\cite{SMT02})
as well as the composition and properties of meteoritic, cometary, and
interplanetary dust together with their laboratory analogues (e.g.,
Chihara et al.~\cite{Cea02}; Mutschke et al.~\cite{MPFD02}) form a
basis for theoretical investigations.

On the other hand, the increase of computer power and new numerical
methods stimulate the development of more sophisticated hydrodynamical
models (e.g., Kley, D'Angelo, \& Henning~\cite{KDH01}; D'Angelo, Henning, \& 
Kley~\cite{AHK02}; Struck, Cohanim, \& Willson~\cite{SCW02}).
Many of such simulations need an accurate treatment of the energy
transport within the dusty medium (see e.g., Klahr, Henning, \&
Kley~\cite{KHK99}; Niccolini, Woitke, \& Lopez~\cite{nwl03}), which requires 
a detailed description of the radiative properties of matter. Consequently, 
the adopted opacity model is an important issue.

In this paper, we deal with physical conditions typical for
protostellar nebulae and protoplanetary discs around low-mass
young stellar objects. Virtually
everywhere within the medium dust grains are the main opacity
source, as they absorb radiation much more efficiently compared to
the gas and because the temperature in these regions is low 
enough to prevent their destruction. However, for hotter 
domains ($T \ga 1\,500$~K), where even the most stable dust 
materials cannot survive, it is necessary to take absorption and 
scattering due to molecular species into account.

Recently, several extensive models describing the properties and
evolution of dust grains in protostellar cores and protoplanetary
discs were proposed by Henning \& Stognienko~(\cite{HS96}),
Schmitt, Henning, \& Mucha~(\cite{Sea97}), and
Gail~(\cite{G01},~\cite{G02}).

Henning \& Stognienko~(\cite{HS96}) studied the influence of
particle aggregation on the dust opacity in the early evolutionary
phases of protoplanetary discs. They concluded that distribution
of iron within the particles affects their optical properties in a
great respect. Schmitt et al.~(\cite{Sea97}) for the first time
investigated collisional coagulation of dust grains in
protostellar and protoplanetary accretion discs coupled with
hydrodynamical evolution of these objects. They reported
significant alteration of the thermal disc structure caused by the
modification of the opacity due to dust growth.
Gail~(\cite{G01,G02}) considered annealing and combustion
processes leading to the destruction of silicate and carbon dust
grains consistently with the evolution of a steady-state accretion
disc. They found that the modification of the dust composition in
the inner regions due to these processes and its consequent
transport toward outer disc domains affect the opacity and,
eventually, the entire disc structure.

A number of papers deal with the computation of Rosseland or/and
Planck mean gas opacities in atmospheres of cool stars, protostars,
and stellar winds. Alexander \& Ferguson~(\cite{AF94}) computed a set
of opacity tables for temperatures between $700$~K and $12\,500$~K for
several compositions. They considered a condensation
sequence\footnote{which is not to be confused with a formation
sequence (see also Woitke~\cite{woi2000})} of refractory materials
based on chemical equilibrium calculations and took into account
absorption and scattering properties of these solids as well as
various gas species. Note that the actual formation process is not
modelled by such calculations. Helling et al.~(\cite{H00})
calculated gas mean opacities for wide ranges of density,
temperature and various chemical compositions based on up-to-date
spectral line lists of the Copenhagen SCAN data base and studied
the importance of the molecular opacity for the dynamics of the
stellar winds of cool pulsating stars. An extension of this work
will be used to construct our opacity model for protoplanetary
accretion discs.

In Appendix~A, we give a brief overview of the most common opacity models and
studies and where they have been applied. It can be clearly seen 
that there is a lack of papers which focus on calculations of 
both Rosseland and Planck mean opacities of grain and gas species 
for temperatures between several K and few thousands K in a wide 
range of densities based on both the best estimates of the dust 
composition and properties and recent improvements in molecular 
line lists.

The first goal of this paper is to define such a model. In addition,
we study how Rosseland and Planck mean opacities depend on the
properties of dust grains and compare them with other opacity tables.
Furthermore, we investigate how different opacity models affect the
hydrodynamical structure of accretion discs. Our opacity model is
freely available in the Internet: {\it
  http://www.astro.uni-jena.de/Laboratory/labindex.html}.

The paper is organised as follows. We introduce the opacity model
in Sect.~\ref{model}. The influence of the grain properties on 
resulting Rosseland mean
opacities is described in Sect.~\ref{gp}. The Rosseland and Planck
mean opacities are compared to other recent opacity tables in
Sect.~\ref{com}. We study how different opacity tables affect
the hydrodynamical structure of active accretion discs in
Sect.~\ref{struc}. A summary follows in Sect.~\ref{sum}. Finally, 
an overview of the opacity models cited in the paper is given in 
Appendix~A.

%======================================================================
% II. Model:
%======================================================================
\section{The opacity model}
\label{model}

%======================================================================
% II.1 Dust opacities:
%======================================================================
\subsection{Dust opacities}

In order to calculate frequency-dependent, Rosseland and Planck
mean opacities of dust grains, we partly follow the schemes,
proposed in Pollack et al.~(\cite{Pea94}, hereafter PHB), Henning
\& Stognienko (\cite{HS96}, HS), and Voshchinnikov \&
Mathis~(\cite{VM99}, VM).

%======================================================================
% II.1.1 Dust composition, size distribution etc.:
%======================================================================
\subsubsection{Composition model and grain sizes}

%Christiane
In this paper we follow the dust composition model for accretion
discs by Pollack et al.~(\cite{Pea94}; Sect. 2.3) which is based on an
analysis of a wide range of theoretical, laboratory, and observational dust
data. This composite model has been frequently used, for instance, in the
evolutionary modelling of accretion discs around young stellar
objects or estimates on their mass from millimetre observations
(e.g., Greaves et al.~\cite{Gea98}; Jura \& Werner~\cite{JW99}; D'Alessio, 
Calvet, \& Hartmann~\cite{DCH01}).

%Dmitry
%We use the solar elemental abundances for all elements from the
%compilation of Anders \& Grevesse~(\cite{AG89}) and for carbon
%from Grevesse et al.~(\cite{Gea91}). Essentially the same dust
%composition is adopted as described in PHB. This model is defined
%on the basis of an accurate analysis of a wide range of 
%theoretical, laboratory, and observational dust data. Given that, 
%it has been used in the modelling of the evolution of accretion 
%discs around young stellar objects or estimations on their mass 
%from millimeter observations (e.g., D'Alessio, Calvet, \& 
%Hartmann~\cite{DCH01}; Greaves et al.~\cite{Gea98}; 
%Jura \& Werner~\cite{JW99}).
  
The main dust constituents include amorphous pyroxene ([Fe,
Mg]SiO$_3$), olivine ([Fe,Mg]$_2$SiO$_4$), volatile and refractory
organics (CHON material), amorphous water ice, troilite (FeS), and
iron\footnote{See: {\it http://www.astro.uni-jena.de/Laboratory/labindex.html}}. 
Following HS, we vary the relative iron content in the silicates
considering ``iron-rich'' (IRS) silicates with Fe/(Fe+Mg)=0.4,
``normal'' silicates (NRM) with Fe/(Fe+Mg)=0.3, and ``iron-poor''
(IPS) silicates with Fe/(Fe+Mg)=0. However, the absolute amount of
metallic iron in all these models is kept constant, which leads
to the absence of solid iron in the first case and enhanced mass
fraction of Fe in the third case. Such a variety of silicate models
allows us to study the influence of iron content within the grain
constituents on the extinction properties of dust. Another reason is
that the exact mineralogical composition of the silicates in the
protostellar clouds and protoplanetary discs is poorly constrained and
can be different for various environments. 
Compared to HS, we re-estimated the absolute abundances of the 
silicates (Fe$_x$Mg$_{1-x}$SiO$_3$, Fe$_{2x}$Mg$_{2x-2}$SiO$_4$), iron, 
and troilite (FeS) in the cases of the ``iron-poor'' and ``iron-rich''
models from the Fe-stocheometric fractions keeping constant the total
amount of iron. For the mass fractions of all dust
constituents and their densities we follow Table~$2$ in PHB. Note 
the difference between the iron mass fractions in the different dust 
models.

The sublimation temperatures of the grain constituents are adopted
from PHB (see Table~$3$ therein). We suppose that destruction of
dust materials occurs in a narrow range of temperatures ($\sim
10-30$~K). Given that the evaporation of the silicates and iron
happens at approximately the same conditions, we do not
distinguish between their evaporation temperatures and assume
that they evaporate in one wide temperature range, $\sim
100$~K. Thus, we account for six principal temperature regions:
\begin{enumerate}
\item $T\la 155$~K -- all dust constituents are present;

\item $\sim 165 \mathrm{K}< T\la 270$~K -- no ice;

\item $\sim 280 \mathrm{K}< T\la 410$~K -- no ice and volatile
organics;

\item $\sim 440 \mathrm{K}< T\la 675$~K -- silicates, iron, and
troilite are present;

\item $\sim 685 \mathrm{K}< T\la 1\,500$~K -- silicates and iron
are present;

\item $\sim 1\,500 \mathrm{K}< T\la 10\,000$~K -- gas-dominated opacities;

\end{enumerate}

We take into account the dependence of the evaporation temperatures of
ice, silicates, and iron on gas density (the values shown above are
given for a density $\sim 10^{-10}$~g\,cm$^{-3}$). Note that if one
considers a condensation sequence of these materials (hysteresis
behaviour), it results in higher values of the condensation
temperatures compared to the evaporation temperatures.

We assume that for the fifth temperature region the absolute
amount of solid iron raises due to the destruction of troilite.
The corresponding mass fractions of metallic iron are $6.15 \times
10^{-4}$, $2.42 \times 10^{-4}$, and $1.29 \times 10^{-3}$ for the
NRM, IRS, and IPS silicate models, respectively.

As for the size distribution, we apply a modified MRN (Mathis et
al.~\cite{MRN}) function suggested by Pollack, McKay, \&
Christofferson~(\cite{P85}). The modification consists of the
inclusion of large ($0.5 \mu$m -- $5 \mu$m) dust grains. Such
particle growth is expected to proceed efficiently at the early
phases of the protoplanetary disc evolution due to the coagulation
of small dust grains (see, for instance, Mizuno, Markiewicz, \&
Voelk~\cite{MMV88}). We do not consider other size distributions
since the overall effect of the particle sizes on the dust opacity
is well studied (e.g., Pollack et al.~\cite{P85}; Beckwith,
Henning, \& Nakagawa~\cite{BHN00}). It further allows us to
compare directly our results with other works.

%======================================================================
% II.1.2 Topology of dust grains and their shapes:
%======================================================================
\subsubsection{Grain structure and topology}

It becomes evident from theoretical investigations and
laboratory experiments that the dust agglomeration is an efficient
process in dense and relatively cold environments, like
protostellar cores or protoplanetary discs (e.g., Kesselman~\cite{K80}; 
Nuth \& Berg~\cite{NB94}; Ossenkopf \& Henning~\cite{OH94}; Kempf, Pfalzner, 
\& Henning~\cite{KPH99}; Wurm \& Blum~\cite{WB98},~\cite{WB00}; Blum et
al.~\cite{Bea02}). 
Agglomeration leads to the formation of irregular particles 
consisting of hundreds or thousands of tiny subgrains. 
Usually, dust aggregates of two extreme kinds are considered, depending on the
assumed coagulation processes, namely, PCA (particle-cluster
aggregation) and CCA (cluster-cluster aggregation). As the laboratory
and theoretical studies reveal, the PCA aggregates are sphere-like particles having a
fractal dimension of about $3$. They have a compact ``core'' and a
more rarefied ``mantle''. The CCA process results in the formation of
very filamentary grains with complicated structure. They have fractal
dimension of roughly $2$ (Stognienko, Henning, \&
Ossenkopf~\cite{SHO95}).
%It is 
%worthwhile to mention that the interplanetary dust particles have 
%a structure similar to that of the laboratory analogs, the PCA 
%(Rietmeijer \& Nuth~\cite{RN00}).

During the evolution of parent objects, like molecular clouds or protostellar discs, chemical and physical processes can further modify the properties of dust grains. For instance, accretion of volatile materials on dust
surfaces and their subsequent chemical processing are efficient in
outer regions of protoplanetary discs and in protostellar clouds
(e.g. Greenberg~\cite{G67}; Brown, Charnley, \& Millar~\cite{BCM88};   
Hartquist \& Williams~\cite{HW90}; Hasegawa \& 
Herbst~\cite{HH93}; Willacy, Rawlings, \& Williams~\cite{WRW94};
Aikawa et al.~\cite{Aea99}). 
This results in well-defined ``core-mantle''
or, more probably, ``onion-like'' grain structure. In protostellar
discs, dust can be transported by the accretion flow toward hotter
regions, where their volatile mantle materials evaporate, and
sputtering, annealing, combustion, and crystallisation processes
may change the structure, composition, and sizes of the grains
(Gail~\cite{G01,G02}). Therefore, it seems obvious that the real
astronomical grains must have a very complicated structure and
topology.

Unfortunately, modern computational methods and facilities allow
only the consideration of somewhat simplified (but still
reasonable) types of dust grains. In the present study, we focus
on the following particle types:
\begin{enumerate}
\item Homogeneous and composite aggregates;
\item Homogeneous, composite, and porous composite spherical
      particles;
\item Multishell and porous multishell spherical particles.
\end{enumerate}

An aggregate dust particle is assumed to be a cluster of small
spherical subgrains sticking together. A particle is called
``homogeneous'' if it consists of only one dust component. On the
contrary, ``composite'' means that a particle incorporates a fine
mixture of various materials (heterogeneous particle). In 
addition to these two extreme cases of the chemical dust 
structure, we consider a model of ``multishell'' grains, where 
each particle includes all constituents distributed within 
concentric spherical shells. To study the influence of the 
porosity on the extinction properties of dust grains, we fill 
composite and multishell spherical particles with vacuum. It is 
reasonable to assume that the optical behaviour of these porous 
multishell and porous composite particles may resemble that of 
more realistic kinds of dust grains.
%======================================================================
% Table N1 - Mass frac. / density in ADS
%======================================================================
\begin{table*}
   \caption[]{Mass fractions $f_j$ and densities $\rho_j$ of dust constituents}
   \label{rmf}
   \begin{tabular}{llll}
   \noalign{\smallskip}
\hline
   \multicolumn{1}{l}{ Material } &
   \multicolumn{1}{l}{ NRM } &
   \multicolumn{1}{l}{ IRS } &
   \multicolumn{1}{l}{ IPS }\\
   \multicolumn{1}{l}{ (grain species $j$) } &
   \multicolumn{1}{l}{ $f_j$ \hspace*{1cm} $\rho_j$} &
   \multicolumn{1}{l}{ $f_j$ \hspace*{1cm} $\rho_j$ } &
   \multicolumn{1}{l}{ $f_j$ \hspace*{1cm} $\rho_j$}\\
%   \multicolumn{1}{l}{ } &
%   \multicolumn{1}{l}{ } &
%   \multicolumn{1}{l}{ } &
%   \multicolumn{1}{l}{ }\\
\hline
   Olivine   & $2.64\,10^{-3}$ ($3.49$ g\,cm$^{-3}$) & $3.84\,10^{-3}$ ($3.59$ g\,cm$^{-3}$)
             & $6.30\,10^{-4}$ ($3.20$ g\,cm$^{-3}$)\\
%\hline
   Iron      & $1.26\,10^{-4}$ ($7.87$  g\,cm$^{-3}$) & --
             & $7.97\,10^{-4}$ ($7.87$  g\,cm$^{-3}$)\\
%\hline
   Pyroxene  & $7.70\,10^{-4}$ ($3.40$  g\,cm$^{-3}$) & $4.44\,10^{-5}$ ($3.42$ g\,cm$^{-3}$)
             & $1.91\,^{-3}$ ($3.20$  g\,cm$^{-3}$)\\
%\hline
   Troilite  & $7.68\,10^{-4}$ ($4.83$  g\,cm$^{-3}$) & $3.80\,10^{-4}$ ($4.83$ g\,cm$^{-3}$)
             & $7.68\,10^{-4}$ ($4.83$  g\,cm$^{-3}$) \\
%\hline
   Refractory&                 &                   &\\
   organics  & $3.53\,10^{-3}$ ($1.50$  g\,cm$^{-3}$) & $3.53\,10^{-3}$ ($1.50$ g\,cm$^{-3}$)
             & $3.53\,10^{-3}$ ($1.50$  g\,cm$^{-3}$) \\
%\hline
   Volatile  &                   &                   &\\
   organics  & $6.02\,10^{-4}$ ($1.00$  g\,cm$^{-3}$) & $6.02\,10^{-4}$ ($1.00$ g\,cm$^{-3}$)
             & $6.02\,10^{-4}$ ($1.00$  g\,cm$^{-3}$) \\
%\hline
   Water ice & $5.55\,10^{-3}$ ($0.92$  g\,cm$^{-3}$) & $5.55\,10^{-3}$ ($0.92$ g\,cm$^{-3}$)
             & $5.55\,10^{-3}$ ($0.92$  g\,cm$^{-3}$) \\
\hline
\end{tabular}
\end{table*}

%======================================================================
% II.1.3 Computational approaches:
%======================================================================
\subsubsection{Computational methods}
\label{methods}

The aggregate model and the numerical method to compute the
optical properties of coagulated particles are adopted from HS.
The aggregates are assumed to be in the form of PCA (50\%) and CCA
(50\%) particles consisting of $0.01 \mu$m spherical subgrains.
The spectral representation of inhomogeneous media
(Bergman~\cite{B78}) and the discrete multipole method (DMM) are
elaborated to calculate effective dielectric functions of
aggregates (Stognienko et al.~\cite{SHO95}). At first, we use the
DMM to calculate the spectral function of the aggregated particles
of a special topology, when their subgrains touch each other only
at one point. Then, we account for the interaction strength
between the subgrains (percolation), which varies with the size
and type of the aggregates, by an analytical expression (see
Equation~$13$ in HS). Finally, the effective dielectric functions
of the aggregates are obtained by the spectral representation (HS,
Eq.~$5$). In the case of the composite aggregates, we compute the
optical constants of the composite material by the Bruggeman
effective medium theory (Bruggeman~\cite{B35}), generalised to
many components by Ossenkopf~(\cite{O91}). The optical properties
of the dust aggregates are calculated with the usual Mie theory.
It should be noted that this numerical approach is valid in the
static limit only, which means that the scale of inhomogeneities
within the particles must be small compared to the wavelength.
Given the $0.01 \mu$m size of the subgrains and the shortest
wavelength of $0.1 \mu$m we considered, this condition is
fulfilled.

We use the numerical approach of VM to model (porous)
composite and (porous) multishell dust particles. In this method, the
composite grains are represented as spheres with many
concentric shells, where each shell includes several layers of
randomly distributed dust materials. The multishell grains
are modelled exactly as the composite ones but each shell includes
only one layer of a dust constituent. Then a generalised multilayered 
Mie theory can be applied to calculate their optical properties. As it 
has been shown by VM, a convergence in the optical behaviour of the
multishell particles is achieved if the number of shells exceeds
$3$ and dust materials within shells are randomly ordered. In our
calculations, we found that this number must be at least $20$,
because the highly absorbing material troilite is
used, which induces interference within the shells and
prevents fast convergence. Thus, in our case, a typical composite
grain is modelled as a spherical particle with about hundred
shells. On the contrary, a multishell grain is represented by a
spherical particle with only a few shells.

We modify somewhat the dust model for the case of multishell and
composite spherical particles. We mixed the silicates and iron into 
one material with the Bruggeman mixing rule. A similar mixture of 
silicates, sulphides, and metals
(GEMS, Glass with Embedded Metals and Sulphur) is found to be a
common component of interplanetary dust particles (Rietmeijer \&
Nuth~\cite{RN00}). However, the main reason for this change is the
convergence failure of the applied numerical method for the case
of multishell grains with iron layers. This is due to numerical
uncertainty which arises in the calculation procedure for the Mie
scattering coefficients in the case of highly absorbing materials,
like iron (for further explanation, see Gurvich, Shiloah, \&
Kleiman~\cite{GSK01}).

We assume that each dust component has a total volume fraction in
a particle according to its mass fraction and density, as
specified in Table~\ref{rmf}. For example, for the first
temperature region and in the case of IPS silicate mineralogy, the
mixture of iron and silicates occupies $8.9$\%, troilite --
$1.6$\%, refractory organics -- $23.4$\%, volatile organics --
$6$\%, and water ice -- $60$\% of the entire particle volume,
respectively. These values are similar in the case of NRM and IRS
models. Thus, if the temperature is low, organics and ice are the
dominant components of the dust grains.

Unlike to the case of the composite particles, in the case of
multishell spherical grains it is assumed that the distribution of dust
materials is not random but follows their evaporation sequence.
Thus, for the first temperature region in the protoplanetary disc,
multishell spherical grains consist of a refractory core made of
a mixture of silicates and iron and subsequent shells of
troilite, refractory organics, volatile organics, and water ice.
For higher temperature ranges, the number of shells is smaller
since some materials are evaporated. In total, the number of shells in
the case of multishell spherical particles varies from $2$ to $5$.
For the fifth temperature range ($T>700$~K), where all troilite is
converted to solid iron, we let this iron form an additional
layer on the grain surface. It is an extreme and probably physically
unjustified case, but this allows us to study the influence of the
formation of a highly absorbing surface layer on resulting opacities.

The porosity of particles is treated in a simple manner by the
addition of $50\%$ of vacuum (by volume) inside. For the case of
the porous composite spheres, we consider vacuum to be one of the
grain constituents, which forms additional "empty" layers. In the
case of the multishell grain model, we mix each shell with vacuum in
the same way which is applied to create the composite spherical
particles. That is, we subdivide each individual shell in many
layers and fill some of them with vacuum according to the
requested porosity degree.

With the two computational approaches described above, we calculated
the ensemble averaged absorption and scattering cross section as well
as albedo and the mean cosine of the scattering angle for all kinds of
dust grains. Applying Eqs.~$1$-$5$ from Pollack et al.~(\cite{P85};
see Table~\ref{rmf} of the present paper), the dust
monochromatic opacity and, consequently, the Rosseland and Planck mean
opacities were obtained for temperatures below roughly $1\,500$~K and
a density range between $10^{-18}$~g\,cm$^{-3}$ and
$10^{-7}$~g\,cm$^{-3}$. A convenient analytical representation of the
Rosseland and Planck mean opacities for every temperature region is
provided as a $5$-order polynomial fit. This representation allows to
calculate the Rosseland and Planck mean dust opacities accurately
($\sim 1$\%) and quickly for any given temperature and density values
within the model applicability range, which is important for
computationally extensive hydrodynamical simulations. The
corresponding fit coefficients can be found in the 
code\footnote{http://www.astro.uni-jena.de/Laboratory/labindex.html}.

%======================================================================
% II.3 Gas opacities:
%======================================================================
\subsection{Gas opacity}

The opacity of the very inner part of the protoplanetary accretion
disc is dominated by various gaseous species. Here, the
temperature is too high for dust to be present.

Compared to the calculation of dust opacities, the calculation of
accurate Rosseland and Planck mean gas opacities is more
challenging due to the large variation in frequency, temperature,
and density of the absorption coefficient of numerous molecules,
atoms, and ions. In addition, the body of data to be handled
easily amounts to several millions of absorption lines per
molecule. 

Missing data for absorption lines are critical for the calculation
of Rosseland mean gas opacity since it is dominated by transparent
spectral regions due to the harmonic nature of the averaging process.
Therefore, each Rosseland mean is always only a lower limit
of the correct value. The opposite is true for the case of the Planck
mean opacity -- missing data for weak lines or bands cause an
overestimation of the strong lines. Therefore, a Planck mean is
always an upper limit of the case of ideally complete data.

The dust opacity model for protoplanetary accretion discs outlined in
the previous sections is supplied by a new table of gas opacities
assembled on the basis of Helling~(\cite{H_PhD99}; Copenhagen SCAN
data base) and Schnabel~(\cite{S_MT01}; HITRAN data base).  The
gas opacity model is outlined in Helling et al.~(\cite{H00}) and only
a short summary is given here. The Rosseland and the Planck mean
opacities are calculated from opacity sampled lines lists. The data
for the line absorption coefficients used in Helling et
al.~(\cite{H00}) (CO - Goorvitch \& Chackerian~~\cite{CO}; TiO -
J{\o}rgensen~\cite{TiO}; SiO - Langhoff \& Bauschilder~\cite{SiO};
H$_2$O - J{\o}rgensen \& Jensen~\cite{H2O}; CH - J{\o}rgensen et
al.~\cite{CH}; CN, C$_2$ - J{\o}rgensen \& Larsson~\cite{CN}; C$_3$ -
J{\o}rgensen~\cite{C3}; HCN, C$_2$H$_2$ - J{\o}rgensen~\cite{C2H2})
were supplemented by data from the HITRAN\,96 database (CH$_4$,
NH$_3$, HNO$_3$, H$_2$CO, CO$_2$, N$_2$O, O$_3$, SO$_2$, NO$_2$,
HO$_2$, H$_2$, O$_2$, NO, OH, N$_2$). The opacity sampling of the
latter was carried out by Schnabel~(\cite{S_MT01}). The set of
continuum opacities and scattering includes continuum absorption from
\ion{H}{i} (Karzas \& Latter~\cite{Hi}), H$^-$ (John~\cite{H-}), H$+$H
(Doyle~\cite{H&H}), H$_2^-$ (Somerville~\cite{H2-}), H$_2^+$
(Mihalas~\cite{H2+}), He$^-$ (Carbon et al.~\cite{He-}), \ion{He}{i},
\ion{C}{i}, \ion{Mg}{i}, \ion{Al}{i}, \ion{Si}{i} (all from
Peach~\cite{P70}) as well as Thompson scattering on free electrons and
Rayleigh scattering.
% for \ion{H}{i} and \ion{He}{i} (Dalgarno~\cite{D62}) 
Collision-induced absorption has been considered for H$_2$-H$_2$ and
H$_2$-He according to Borysow et al.~(\cite{bjz97}).

The number densities of the ions, atoms and molecules are computed
from an updated chemical equilibrium routine, including 14 elements and
155 molecules based on the JANAF table (electronic version of Chase et
al.~\cite{Ch85}; for more detail see Helling et al.~\cite{H00}). The element
abundances are chosen mainly according to Anders \& Grevesse~(\cite{AG89}), but
have been updated for various elements (see Helling et al.~\cite{H00}).

%======================================================================
% Figure:
%======================================================================
\begin{figure*}
  \includegraphics[width=1.0\textwidth]{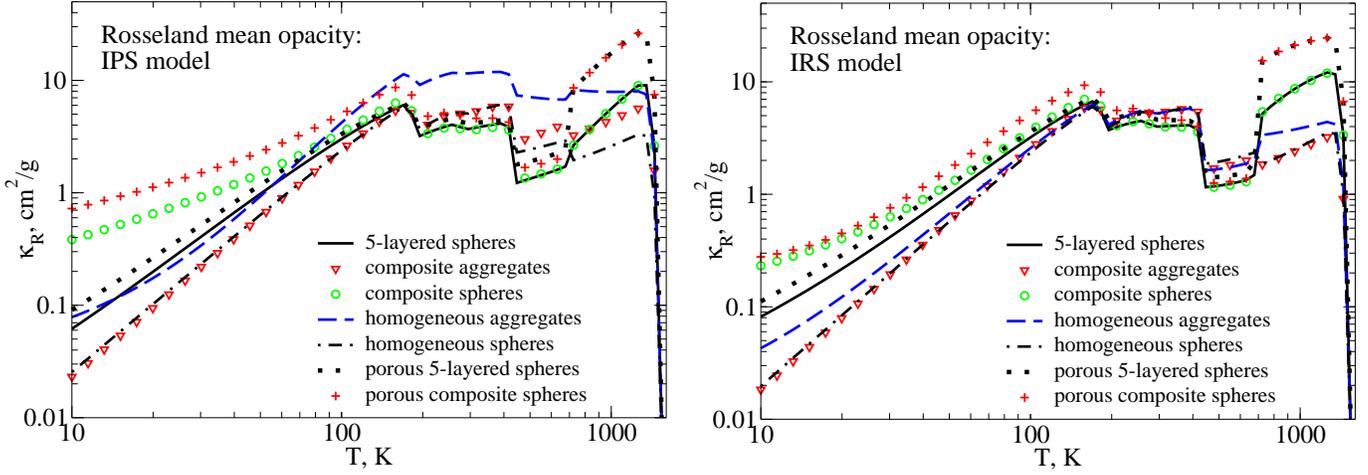}
  \caption{The Rosseland mean opacities in the case of IPS (left panel)
           and IRS (right panel) silicate dust model. The following
           dust particle types are indicated: multishell spheres -- solid line, 
           composite aggregates -- triangles, composite spheres -- circles, 
           homogeneous aggregates -- dashed line, homogeneous spheres -- 
           dot-dashed line, porous multishell grains -- dotted line, and 
           porous composite spheres -- pluses.}
  \label{f1}
\end{figure*}

Using this approach, the Rosseland and Planck mean gas opacities
were computed for temperature ranges between $500$~K and
$10\,000$~K and for gas densities between $\sim
10^{-18}$~g\,cm$^{-3}$ and $\sim 10^{-7}$~g\,cm$^{-3}$. In
contrast to the dust opacity, no simplified analytical expression
can be found for the gas mean opacities because of their sensitive
dependence on temperature and density. Thus, we apply a
second-order interpolation in order to calculate the gas opacities
for any given temperature and density value from tabulated values.

%======================================================================
% II.4 Rosseland and Planck mean opacities:
%======================================================================
\subsection{Opacity table}

In order to assemble the opacity table, we take into account either
only dust opacity data for low temperatures, $T \la 1\,500$~K, or only
gas opacity data for higher temperatures. As has been shown by many
authors (e.g., Lenzuni, Gail, \& Henning~\cite{L95}), it is an
accurate approach because dust dominates the absorption properties of
matter whenever it is present. It has been shown that already 30\% 
of total SiO content in the solid phase is enough to exceed
even the Planck mean gas opacity (Helling~\cite{H_PhD99}). This
fraction will decrease with increasing impurity of the grains.
However, for the dust-to-gas transitional region ($T$ from $\sim
1\,400$~K to $1\,600$~K), where the last dust grain population
evaporates, it is necessary to calculate the opacity of dust and gas
simultaneously. In this narrow temperature range the resulting
Rosseland and Planck mean opacities are certainly going down by few
orders of magnitude, so one can apply a simple linear interpolation to
estimate their values with a good accuracy.  Still, we note that our
table may give only approximate opacity data for that
temperature range.

%======================================================================
% III. Results:
%======================================================================
\section{Results}
\label{res}

In this section, we present the opacities of all types of dust 
grains and discuss the differences between them. Second, we 
compare the Rosseland and Planck mean opacities with other 
recently published opacity models. Finally, we study the 
influence of the adopted opacity model on the accretion disc 
structure.

%======================================================================
% III.1. Rosseland mean dust opacities:
%======================================================================
\subsection{Opacity and the dust models}
\label{gp}

The Rosseland mean opacities $\kappa_{\mathrm R}$ computed for 
all dust models are presented in Fig.~\ref{f1}. We compared two 
silicate models, namely, the IPS (left panel) and the IRS (right 
panel). Shown are the following dust models: multishell spheres, 
composite aggregates, composite spheres, homogeneous aggregates, 
homogeneous spheres, porous multishell grains, and porous composite spheres.

The most prominent trends in behaviour of the Rosseland mean dust 
opacities can be summarised as follows:
\begin{enumerate}

\item There is a significant difference in the calculated dust opacity 
      values between the aggregates, (porous) composite and (porous) multishell spherical 
      particles for the first ($T \la 150$~K) and fifth ($T \ga 700$~K) temperature 
      regions;

\item For intermediate temperatures, $150~\mathrm{K} \la T \la 700$~K, the resulting 
      Rosseland mean opacities do not show profound variations with the applied dust 
      models, except for the case of the IPS homogeneous aggregates;

\item The discrepancy between opacity curves are smaller for the case of the IRS
      silicate model compared to the IPS;
    
\item Addition of vacuum inside compact composite and multishell spherical particles 
      leads to significant modification of their opacities.

\end{enumerate}
% empty sentence:
%We discuss all these points in detail below.

In the first temperature region, organics and water ice are 
the dominant dust materials according to our compositional model (see 
Table~\ref{rmf}). At that temperatures ($T \la 150$~K), the main 
contribution to the Rosseland mean opacity comes from
monochromatic opacities in the wavelength range about $30 \mu$m -- $400 \mu$m. 
In this spectral range, iron and troilite have higher values of 
the real ($n_\lambda$) and imaginary ($k_\lambda$) indices of 
refraction compared to the other materials considered in our 
model. For instance, at $\lambda = 100 \mu$m, iron has 
$n^\mathrm{Fe}_{100 \mathrm{{\mu}m}} = 95.02$, 
$k^\mathrm{Fe}_{100 \mathrm{{\mu}m}} = 181.95$, troilite has 
$n^\mathrm{FeS}_{100 \mathrm{{\mu}m}} = 8.5$, 
$k^\mathrm{FeS}_{100 \mathrm{{\mu}m}} = 0.73$, whereas organics and 
water ice have $n^{Org}_{100 \mathrm{{\mu}m}} = 2.14$, 
$k^{Org}_{100 \mathrm{{\mu}m}} = 0.15$ and $n^{Ice}_{100 \mathrm{{\mu}m}} 
= 1.82$, $k^{Ice}_{100 \mathrm{{\mu}m}} = 0.05$, respectively. 
Furthermore, for troilite and especially for iron these values 
rapidly increase with wavelength. Thus, the resulting optical 
properties of a dust grain become sensitive to the absolute 
amount of Fe and FeS and its distribution inside the particle. In 
such a case, different theories to calculate optical properties 
of an aggregate particle may give different results and should be 
used with caution (see Stognienko et al.~\cite{SHO95}, 
Fig.~6,~8). 

We considered aggregated particles of two kinds, namely, 
composite and homogeneous aggregates. The optical constants of 
the composite material do not show a peculiar behaviour at long 
wavelengths. Moreover, the refractive index of this composite 
changes only little if one is switching from the IPS to the IRS 
silicate model. One reason is that the amount of solid Fe is small 
compared to other constituent materials. Another reason is that 
the total iron abundance is kept constant in all compositional 
models. Hence the resulting optical properties of the composite 
aggregates are not very sensitive to the actual topology of the 
particles and adopted silicate model. Indeed, as it can be 
clearly seen in the Figure, the dust opacity values of the 
composite aggregates (triangles) are close to the 
$\kappa_{\mathrm R}$ of the homogeneous spheres (dot-dashed 
line). The maximum deviation of these opacity curves is achieved 
in the case of the IPS silicates for temperatures higher than 
$\sim 700$~K. Here, the absolute amount of metallic iron is 
increased due to conversion of FeS to Fe. In addition, the 
Rosseland mean opacities of the composite aggregates do not 
depend much on the adopted silicate model (compare triangles on 
the left and right panel). 

%======================================================================
% Figure N2:
%======================================================================
\begin{figure*}
  \includegraphics[width=\textwidth]{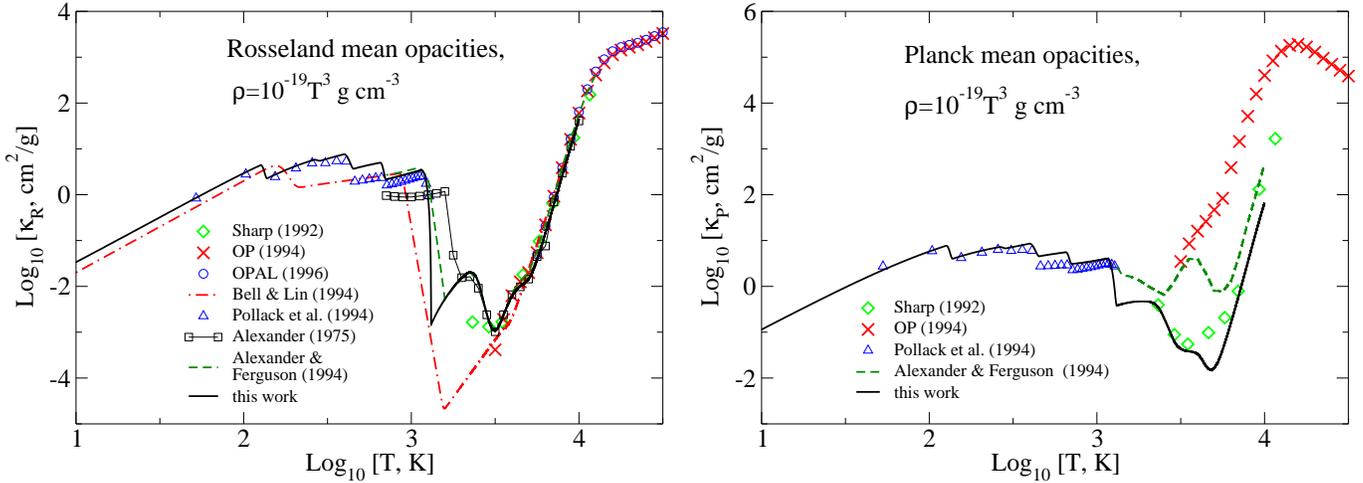}
  \caption{The Rosseland (left panel) and Planck (right panel) mean
           opacities calculated for the whole temperature range
           considered ($T \in [10, 10^5]$~K)
           and gas density scales as $10^{-19} \times T^3$~g\,cm$^{-3}$
           (C/O$=0.43$).
           Depicted are the mean opacities composed of the NRM composite
           aggregates in the low temperature range and the gas opacity
           for the higher temperatures (solid line). In comparison, opacity
           tables of the OP project~(\cite{S94}, crosses), OPAL
           project~(\cite{OPAL96}, circles), Bell \& Lin~(\cite{BL94},
           dot-dashed line), Pollack et al.~(\cite{Pea94}, open triangles),
           Alexander~(\cite{A75}, line with open squares), Alexander
           \& Ferguson~(\cite{AF94}, dashed line), and Sharp~(\cite{Sharp92}, 
           diamonds)} are shown.
  \label{f2}
\end{figure*}

On the contrary, the $\kappa_{\mathrm R}$ values for the 
homogeneous aggregates do demonstrate a strong variations with 
the compositional model. In the case of the "iron-rich" silicates 
(right panel) the Rosseland mean values of the homogeneous 
aggregates (dashed line) lie much closer to the $\kappa_{\mathrm 
R}$ of the composite aggregates (triangles) then for the 
"iron-poor" compositional model (left panel). As it has been 
shown already by HS, this is caused by the presence of bare iron 
aggregates in the case of the IPS homogeneous aggregate model. 
Due to the extremely high absorptivity induced by the strong 
interactions between individual aggregate sub-grains, the optical 
properties of such iron clusters determine the overall behaviour 
of the resulting opacities. Note, that in the case of the 
"iron-rich" silicate composition, all iron is locked inside 
silicates and the absolute amount of troilite is reduced by a factor 
of two compared to the IPS and NRM models. In the absence of a 
population of highly absorbing grains, the Rosseland mean values 
of the homogeneous aggregates are close to that of the composite 
aggregates. However, at temperatures higher than $\sim 700$~K, 
troilite is converted to solid iron which increases the 
$\kappa_{\mathrm R}$ values of the homogeneous aggregates in respect to 
that of the composites (compare dashed lines with triangles on the right 
panel of Fig.~\ref{f1}).

The situation is different for the case of the (porous) composite 
and (porous) multishell spherical particles. As it was mentioned 
in Sect.~\ref{methods}, we changed the compositional model in this case
and locked all solid iron in silicates using the Bruggeman rule of the 
effective medium theory. Thus, 
the only material with a high absorptivity at long wavelengths 
which remains in all compositional models is troilite. The 
metallic iron is another highly-absorbing dust component, but it 
is present in the fifth temperature region only, at $T>700$~K.

Voshchinnikov \& Mathis~(\cite{VM99}) have shown that 
conventional EMTs are rather approximate in the case of small 
composite spherical particles, $x=2\pi a / \lambda \ll 1$, when 
one of the grain constituents has a large refractive index (see 
Fig.~3b therein). Given a typical size $a=0.1 \mu$m of our dust 
grains, a typical wavelength of $\lambda=100 \mu$m for the first 
temperature region, and the high refractive index of troilite at that 
wavelength, this condition is fulfilled. Thus, we adopted the 
approach of VM to model a composite grain as a spherical particle 
with many concentric layers and applied it for all 
temperature regions. Note that there is always an interference 
between the layers, which makes the optical properties of such 
composite particles to be different from that of composite grains 
with well mixed dust constituents. This is especially true if one 
of the dust components has a high absorptivity. In our 
compositional model for layered spherical particles, this is 
troilite in the first temperature region while in the fifth region it is 
iron. Therefore, we may expect to see the difference between the 
Rosseland mean opacity values of the composite spheres and 
composite aggregates, particularly for $T\la 150$~K and $T>700$~K.

As it can be seen in Fig.~\ref{f1}, indeed the $\kappa_{\mathrm R}$ 
values in the case of the composite spherical particles are much 
higher than for the composite aggregates. For instance, at 
$T=10$~K this difference can reach factors of 10 and 20 for the 
IPS and IRS compositions, respectively (compare circles and 
triangles on the left and right panels). For higher temperatures, 
the Rosseland mean opacity curves of the composite spheres and 
aggregates are close to each other till $T \sim 700$~K is 
reached. In this temperature region, we assumed that iron forms 
a layer on the surface of the composite spherical grains. Such a 
layer "screens out" all underlying materials and totally 
dominates the optical behaviour of the entire particle. Due to 
this fact, the dust opacity values of the composite spheres in the 
fifth temperature region is nearly the same for both the IPS and 
IRS compositions.

The multishell spherical particles have a restricted number of 
layers compared to the composite spheres, namely from 2 to 5, 
depending on the temperature region. The troilite layer is assumed to 
be the first layer after the silicate core and thus troilite is 
"hidden" inside. It prevents a strong interference between the 
consequent particle shells as it is the case for the composite 
spheres. Then one may expect that the $\kappa_{\mathrm R}$ values 
of the multishell spheres should be lower than that of the 
composite spherical particles, especially for the IPS model. As 
it can be clearly seen in Fig.~\ref{f1}, this is true for $T\la 150$~K, 
whereas for higher temperatures both opacity curves
almost coincide (compare solid line with circles). Hence, the 
actual distribution of dust constituents within a multilayered 
spherical particle is not that important for the relevant 
Rosseland mean opacities at $T\la 150$~K.

The addition of vacuum inside the compact composite and 
multishell spherical grains leads to a significant increase of 
the corresponding $\kappa_{\mathrm R}$ values for the first and 
fifth temperature regions (compare solid line with dotted line and 
circles with pluses). The first reason is that the density of the 
porous grains becomes lower than the density of the compact 
particles. Second, for the porous spheres the relevant 
extinction efficiencies are higher compared to that of the 
compact spherical particles if some of the dust constituents have 
a particularly high absorptivity, like troilite in the first and 
iron in the fifth temperature region, respectively. This is due 
to the coherence between the particle layers. The interference is 
more intense for the case of the composite sphere since it has 
more concentric layers and a nearly homogeneous distribution of the 
dust constituents from the centre to the surface compared to the 
multishell spherical particle. Note that in the fifth temperature
region, both composite and multishell spheres have a similar
topology, namely, a silicate core covered by iron mantle. Therefore,
it is naturally to expect that they have a similar behaviour of the resulting Rosseland mean opacities.

As it can be clearly seen in Fig.~\ref{f1}, the $\kappa_{\mathrm R}$ values
of the porous composite spheres are higher than that of the compact composite
spherical particles at $T<150$~K and $T>700$~K by a factor of 2 for both
the IPS and IRS compositions (compare pluses and circles). For other temperatures,
the corresponding Rosseland mean curves lie close to each other. The situation
is similar for the case of the multishell and porous multishell spherical particles
(compare solid line with dotted line). As we expected, in the fifth temperature
region the Rosseland mean opacities for the case of the porous composite and 
porous multishell spheres have almost the same values.

%======================================================================
% III.2. The comparison of the opacity models:
%======================================================================
%\subsection{Comparison of the opacity models}
%Christiane
\subsection{Comparison to other studies}
\label{com}

In Fig.~\ref{f2}, the Rosseland (left panel) and Planck (right
panel) mean opacities composed of the NRM composite aggregate for
the low temperature range and gas opacities for the high
temperature range are compared with other models. We plotted these
values for a wide temperature range, $T$ from $\sim 10$~K to
$10^5$~K and for gas densities which scale as $10^{-19} \times
T^3$~g\,cm$^{-3}$. It allows us to make a comparison between the
models in a wide temperature as well as density range
simultaneously.

As it is clearly seen in Fig.~\ref{f2} (left panel), the discrepancy
between the Rosseland mean values provided by various models is
negligible at high temperatures ($T \ga 3\,000$~K). On the contrary,
the Planck mean opacity values differ by a few order of magnitude in
this temperature range (compare curves on the right panel). The reason
is that the Rosseland mean is much less sensitive to differences in the material data than the Planck mean due to the nature of the
averaging process. However, the Planck mean heavily depends on the
adopted values of the band and line strengths which vary for different
line lists and on the adopted chemical equilibrium
constants\footnote{Note that the chemical equilibrium constants,
$K_p$, used by different authors can cause differences in the
resulting opacity values because it affects the number density of
species. The same effect will be caused by the neglect of the
metal ions in the chemical equilibrium calculations (see, e.g.,
Helling et al.~\cite{H00}). It is, however, not very
straightforward to decide which line list is the most correct (for a
discussion see J{\o}rgensen~\cite{ugj03}). We nevertheless dare to
demonstrate the difficulties arising from comparing Planck mean
opacities calculated by different authors because it has -- to our
knowledge -- not been pointed out clearly in the literature. Only
comparisons of Rosseland mean opacities are presented, e.g. by
Alexander \& Ferguson~(\cite{AF94}).}

Therefore, the $\kappa_\mathrm{P}$ of the Opacity Project (OP,
crosses) are much larger than all the other opacity models since it
combines more atomic opacity sources (see e.g. Table~$3.3$ in
Helling~\cite{H_PhD99}). This model does not contain molecules for
temperatures $1\,500$~K -- $5\,000$~K, which become important
absorbers in this temperature range. The difference between the Planck
mean opacity values in the case of the AF model (dashed line) and our
model (solid line) may be caused by different molecular line data
and a different set of chemical species adopted. The same is
true for the $\kappa_\mathrm{P}$ values of Sharp~(\cite{Sharp92},
diamonds), which lay somewhat in between the Planck mean opacities
of our model (solid line) and the values of Alexander \& Ferguson
(dashed line).

For temperatures lower than about $1\,500$~K, dust grains are the main
opacity source. As it has been shown by Pollack et al.~(\cite{Pea94}),
in this case the difference between the Rosseland and Planck mean opacities
computed for the same model is small, $\sim 30$\% (see Fig.~4b therein).
The reason is that both opacities are dominated by continuum absorption
and scattering rather than absorption lines in this case. In what follows, we focus on the low-temperature Rosseland mean opacities only ($T \la 3\,000$~K).

%======================================================================
% Figure N3:
%======================================================================
\begin{figure*}
  \includegraphics[width=1.05\textwidth]{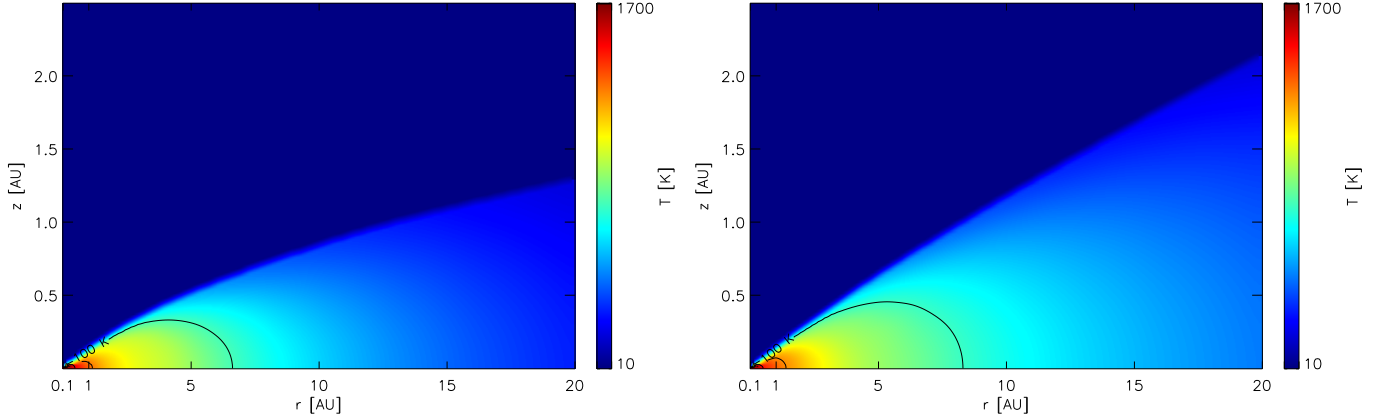}
  \caption{The hydrodynamical structure of the accretion disc derived with
           the BL opacity model (left panel) and in the case of IPS
           homogeneous aggregates (right panel). The solid lines indicate
           temperatures of 1000~K, 500~K, and 100~K, respectively (from the
           left to the right side of the panels).
           }
  \label{f3}
\end{figure*}

The Rosseland mean opacity values $\kappa_\mathrm{R}$ computed by the
model of Bell \& Lin~(\cite{BL94}, BL) strongly deviate from the
$\kappa_\mathrm{R}$ calculated by other models. For example, this
difference can reach a factor of hundred for $T \sim 1\,500$~K --
$1\,800$~K (compare dot-dashed and solid lines in Fig.~\ref{f2}). The reason is that
the BL data are a modification of the old analytical Lin \&
Papaloizou~(\cite{L80}) opacities, which are based on the opacity
tables of Alexander~(\cite{A75}) and Cox \&
Stewart~(\cite{CS70a,CS70b}) supplemented by the data from
Alexander, Augason, \& Johnson~(\cite{AAJ89}) for $T<3\,000$~K.
The model of Bell \& Lin includes dust grains of two types,
namely, homogeneous icy and metallic spherical grains. Since we used the
more advanced dust compositional model of PHB and a different set of
evaporation temperatures, this causes lower $\kappa_\mathrm{R}$ values
in the case of the BL model compared to the other models for $T \la
1\,500$~K. For higher temperatures, $\sim 1\,500$~K -- $3\,000$~K,
this deviation is huge, $\sim 100$ times. As it has been shown by
AF, the reason is that Bell \& Lin truncated monochromatic
opacities of water, which is one of the main absorbers at such 
temperatures, at a too short wavelength in their calculations. The
missing opacity data affect the resulting Rosseland mean opacity.
They do, so far,  not draw any conclusions regarding Planck mean opacities.

In overall, the opacity curves of all other models do not show such a
strong difference between each other (compare triangles, dashed line,
solid line, and line with open squares). Our dust model differs from
the model of Pollack et al. by taking into account an aggregate nature
of cosmic dust grains and a new set of optical constants, but dust
size distribution, composition, and evaporation temperatures are the
same. However, the difference between the Rosseland mean opacity
values of these two model can reach about a factor of two (see also
Fig.~5a in Henning \& Stognienko~\cite{HS96}).

The $\kappa_\mathrm{R}$ values of the opacity model of
Alexander~(\cite{A75}, line with open squares) is lower for the
dust-dominated temperature region ($T \la 1\,500$~K) and higher for
the dust-to-gas transitional region ($1\,500~\mathrm{K} \la T \la
1\,700$~K) compared to our model (solid line) by factors of 5 and 100,
respectively. The reason is, as mentioned above, that we used an
approximation to compute $\kappa_\mathrm{R}$ in that temperature
region, where the last dust grains get evaporated, which is not
a very accurate approach. On the contrary, the opacity model of
Alexander assumes the presence (but no evaporation!) of small
$0.1 \mu$m spherical silicate grains in a rather
approximate way, assuming that all dust is homogeneously condensed
when the gas becomes saturated. Since here a phase-transition takes
place, a supersaturated gas would be needed which results in higher
molecular abundances than AF derive from their equilibrium
consideration (for a discussion see e.g. Woitke \&
Helling~\cite{wh2003}). Moreover, the condensation begins at higher
temperature than the value we assumed for the evaporation of the last
grain constituents because of the dust hysteresis. Finally, the
model of AF neglects the presence of other refractory
materials, like iron, in the dust-dominated temperature range, which
makes the relevant Rosseland mean opacity values lower than provided
by our model.

%Dmitry
%{\bf On the contrary, the model of Alexander \& Ferguson does include
%  condensation of several dust materials, namely, iron, silicate,
%  carbon, SiC out of the gaseous phase into small ellipsoidal grain
%  particles.  Therefore, the corresponding Rosseland mean dust
%  opacities nearly coincide with our values (compare dashed and solid
%  lines, respectively). Surprisingly, the same is true even for the
%  dust-to-gas transitional region, where our opacity model gives
%  rather approximate opacity values.}
%Christiane
On the contrary, the model of AF does
consider several dust materials, namely, iron, silicate, carbon,
SiC assumed to be present as small ellipsoidal grain particles.
Therefore, the corresponding Rosseland mean dust opacities nearly
coincide with our values (compare dashed and solid lines,
respectively). The same is true even for the dust-to-gas
transitional region, where our opacity model gives rather
approximate opacity values.

%======================================================================
% III.3. The disc structure:
%======================================================================
\subsection{Opacity and disc structure}
\label{struc}

We compare the thermodynamical structure of a typical
protoplanetary disc around a low-mass star computed with two
different opacity tables in Fig.~\ref{f3}. The $1+1$D model of an
active steady-state accretion disc of Ilgner~(\cite{I03}) was used
with the following input parameters: $M_{\star}=1M_{\sun}$,
$\dot{M}=10^{-7}M_{\sun}$~yr$^{-1}$, and $\alpha=0.01$. Here,
$M_{\star}$ is the stellar mass, $\dot{M}$ is the mass accretion
rate, and $\alpha$ is the parameter describing the kinematic
viscosity. Note that this model incorporates a star as a
gravitational center only and does not take into account the effect of
the stellar radiation on the disc structure.

The thermal structure shown on the left panel was obtained with
the Rosseland mean opacity table of Bell \& Lin~(\cite{BL94}). On
the right panel, we present the same disc structure but for the
case of the IPS homogeneous aggregate model (IPSHA) supplemented by
the gas mean opacity. We choose these two opacity models as the
overall difference between them is the largest among the different models (compare solid and dash-dotted lines in Fig.~\ref{f2}).

It can be clearly seen that the higher values of the Rosseland mean
opacity in the case of the IPSHA model leads to a hotter and more
extended disc structure. For instance, the vertical scale height
of the disc at $20$~AU is equal to $1.3$~AU for the former and
$2.1$~AU for the latter opacity models, respectively. Consequently, 
there is also a variation of the density structure between the
models, namely, the disc density is higher for the BL opacity model
compared to that of the IPSHA model. The temperature difference is also
prominent. For example, the midplane temperature of $100$~K, which
roughly corresponds to the ice melting point, is reached at
$6.5$~AU for the model of Bell \& Lin, whereas in the case of the
IPSHA model it is at $\approx 8$~AU.

To confirm our findings, we did a similar comparison with another
code. We used a full 2D hydrodynamical code designed to simulate
the interaction of the protoplanetary disc with a protoplanet
(D'Angelo~\cite{DA01}). The parameters of the model were as
follows: $M_{\star}=1M_{\sun}$, $M_{\mathrm d}=0.01M_{\sun}$,
$\nu=10^{15}$~cm$^2$\,s$^{-1}$, $\mu=2.39$, and $\gamma=1.4$,
where $M_{\mathrm d}$ is the total disc mass, $\nu$ is the
kinematic viscosity, $\mu$ is the mean atomic weight of the gas,
and $\gamma$ is the adiabatic exponent.

%======================================================================
% Figure N4:
%======================================================================
\begin{figure}
  \includegraphics[width=0.4\textwidth]{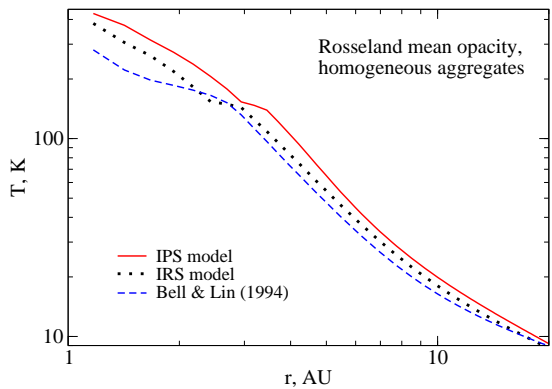}
  \caption{The midplane temperature of the accretion disc obtained
           with the BL opacity model (dashed line) and
           IPS (solid line) and IRS (dotted line) homogeneous aggregate 
           dust models.}
  \label{f4}
\end{figure}

The Bell \& Lin opacity model together with IPS and IRS
homogeneous aggregate models were chosen for the comparison. The
midplane temperature for all three opacity models is shown in
Fig.~\ref{f4}. It can be clearly seen that the difference in the
temperature values between all models can reach about $50$\% for
the disc radii $r_{\mathrm d} \la 2$~AU. Note that it disappears
at larger distances, $r_{\mathrm d} \sim 10$~AU. As expected, in
the case of the BL model the midplane temperature is the lowest
almost everywhere (dashed line), whereas for the IPS homogeneous
aggregates it is the highest (solid line) and the IRS temperature
values lie between them (dotted line). The reason is the same as
for the case of the $1+1$D disc model, namely, lower opacity values of
the BL model compared to both the IPS and IRS opacities and higher
values of IPS opacity in comparison with those of BL and IRS
models. It is interesting that at $T \sim 130$~K ($r \sim 2$~AU) all the
temperature curves are very close to each other. This is due to
the adopted ice evaporation temperatures, which is a little lower
in our case compared to the model of Bell \& Lin. That leads to
nearly the same opacity values for all three models at a
restricted set of temperatures around $\sim 130$~K (compare solid
and dot-dashed lines in Fig.~\ref{f3}). For $r \ga 10$~AU
the temperature curves are close to each other because the corresponding
opacity models have rather similar low-temperature opacity
values.

Thus, we showed that the difference in the Rosseland opacity
tables applied in hydrodynamical calculations leads to 
deviations in the resulting disc structure. As it has been shown
by Markwick et al.~(\cite{Mea02}), the temperature distribution is
a crucial factor for the chemical evolution in the inner parts of
accretion discs. Therefore, proper opacity modelling is an
important issue in order to follow the dynamical and especially
chemical evolution of protoplanetary accretion discs.

%======================================================================
% IV. Summary:
%======================================================================
\section{Summary}
\label{sum}

We have compiled the Rosseland and Planck mean dust and gas
opacities for the temperature range $T \in [5, 10^4]$~K and for gas
densities $\rho \in [10^{-18}, 10^{-7}]$~g\,cm$^{-3}$ which is
appropriate for the conditions in protoplanetary discs. The
absorption and scattering due to dust grains of different
compositions, shapes and topological structures and the
absorption provided by over $30$ atoms, molecules, ions were
taken into consideration. The corresponding well-documented numerical
code together with representative data and figures are electronically
available: {\it http://www.astro.uni-jena.de/Laboratory/labindex.html}.

It has been found that the topological distribution of highly absorbing 
materials, such as iron and troilite, within dust grains 
dominates the relevant optical properties to a large degree. Particularly,
the composite and porous composite spherical grains modelled in a special 
way have remarkably high Rosseland mean opacities at $T\la 150$~K and
$T\ga 700$~K even in comparison with the composite aggregated particles. 
We have shown that at intermediate temperatures the Rosseland mean opacities
of distinct grain models are close to each other. It has been demonstrated that the 
difference between the opacity values of various dust models is smaller in the 
case of the compositional model with a smaller amount of solid iron and troilite.
We found that porous composite and porous multishell spherical particles 
show higher opacity values in comparison with their compact analogues.

We performed a comprehensive comparison of our results with other recent
opacity models. We found a significant difference between the opacity models in the
case of the Planck mean and a good agreement between them for the
case of the Rosseland mean at $T \ga 1\,500$~K, where gas species
are the main opacity sources. For lower temperatures, where
opacities are dominated by dust grains, there is a discrepancy (a
factor of $\sim 3$ at most) in both the Rosseland and Planck mean
values for all considered models.

We demonstrated that differences in the Rosseland mean
opacity values provided by distinct opacity models affect the hydrodynamical 
structure of active steady-state accretion discs. Namely,
higher values of the Rosseland mean opacity lead to a hotter and more 
extended disc structure in the case of $1+1$D and $2$D disc modelling.

%======================================================================
% The acknowledgments:
%======================================================================
\begin{acknowledgements}
DS was supported by the German \emph{Deut\-sche
For\-schungs\-ge\-mein\-schaft, DFG\/} project ``Research Group
Laboratory Astrophysics'' (He 1935/17-1). The work of MI was
supported by the Max Planck Society. We are grateful to D.~Alexander for valuable
comments and discussions. For the calculations of the optical
properties of multishell spherical particles, we used a code by
N.Voshchinnikov ({\it
http://www.astro.spbu.ru/staff/ilin2/ilin.html})
\end{acknowledgements}

%======================================================================
% Appendix:
%======================================================================
\appendix
%======================================================================
% Appendix A:
%======================================================================
\section{An overview of the opacity models}

The goal of this compilation is to provide essential
information about opacity models cited in the text. We show for
which elemental composition they are developed, what kind of
opacities they supply, and in what temperature and density ranges
they work. In addition, key references to the studies, where these
opacity models have been applied, are given and the aims of the
investigations are briefly mentioned.

Primarily, we distinguish between two kinds of opacity models. The
first kind is designed for stellar evolutions, where it
is more convenient to use a special parameter, $R=\rho / T^3_6,
{T}_6 = T / 10^6$~K instead of gas density $\rho$ (see discussion in
Rogers \& Iglesias~\cite{RI92}). Thus, opacity data of such models
are assembled in rectangular tables in R--$T_6$ space. However, in
hydrodynamical simulations of accretion discs, it is more
convenient to have opacity tables composed in $\rho$--$T$
parameter space, for which we refer as to the second type of the
models. The direct conversion of R to $\rho$ in the opacity models
of the first kind leads to trapezoidal tables, where opacity
values for different temperatures have different density intervals. We
mark such models with ``a'' and show the maximum limits of density
and temperature values for them.

In addition, we point out if opacity data are available on-line in
the Internet or via E-mail (``b''). By default, the papers are
supposed to contain the opacity data in a tabular form. Otherwise,
when only opacity plots are available, we label the corresponding
papers with ``c''. The analytical opacity models are marked with 
``d''.

 \begin{table*}
  \caption[]{Opacity models cited in the text}
  \begin{tabular}{llllllllll}
\hline \noalign{\smallskip}
   %% 1st row,
   Model & Composition, & \multicolumn{2}{c}{Absorbers} &
   \multicolumn{2}{c}{Quantities} &
   \multicolumn{2}{c}{Validity} & Usage & Studies \\
\noalign{\smallskip} \cline{2-4} \cline{5-6} \cline{7-8}
\noalign{\smallskip}
   %% 2nd row,
   & Z & dust & gas & $\kappa_{\mathrm R}$ & $\kappa_{\mathrm P}$
   & T, K & $\rho$, g\, cm$^{-3}$ &  & \\
\noalign{\smallskip} \hline \noalign{\smallskip}
   %% 3rd row,
   Cox \& & $10^{-4} \div 1$ & -- & a$^*$ & + & -- &
   $1\,500 \div 10^9$ & $10^{-15} \div 10^{10}$ & Wood &
   Pulsations of \\
   Stewart &  &  &  &  &  &  &  & (\cite{W76}) & luminous\\
   (\cite{CS70a}a,b)$^{\mathrm a}$ &  &  &  &  &  &  &  &  & helium stars\\
\noalign{\smallskip} \hline
 \noalign{\smallskip}
   %% 5th row,
   Alexander & $10^{-3} \div 0.02$ & + & m$^{**}$ & +
   & -- & $700 \div 10^4$ & $10^{-19} \div 10^{-2}$ &
   Bodenheimer & Protostellar\\
   (\cite{A75})$^{\mathrm a}$ &  &  &  &  &  &  &  & et al. (\cite{B90}) &
   collapse\\
\noalign{\smallskip} \hline \noalign{\smallskip}
   %% 6th row,
   Alexander & $10^{-4} \div 2\,10^{-2}$ & + & m  & +
   & -- & $650 \div 10^4$ & $10^{-18} \div 10^{-2}$ &
   Ruden \& & Protostellar \\
   et al. (\cite{A83})$^{\mathrm a}$ &  &  &  &  &  &  &  & Pollack (\cite{RP91}) &
   collapse \\
\noalign{\smallskip} \hline \noalign{\smallskip}
   %% 7th row,
   Pollack et & Solar & + & -- & +$^{\mathrm c}$ & -- & $10 \div 2\,500$ &
   $10^{-14} \div 1$ & Boss (\cite{B88}), & Molecular cloud\\
   al. (\cite{P85}) &  &  &  &  &  &  &  &  &
   fragmentation;\\
   &  &  &  &  &  &  &  & Lunine et & brown dwarf\\
   &  &  &  &  &  &  &  & al. (\cite{L89}), & atmospheres;\\
   &  &  &  &  &  &  &  & Bodenheimer &\\
   &  &  &  &  &  &  &  & et al. (\cite{B90}), & protostellar\\
   &  &  &  &  &  &  &  & Ruden \& & collapse\\
   &  &  &  &  &  &  &  & Pollack (\cite{RP91}) &\\
\noalign{\smallskip} \hline \noalign{\smallskip}
   %% 8th row, 
   Sharp & Solar,  & -- & m & + & + & $2\,10^3 \div 10^4$ & 
   $10^{-10} \div 10^{-3}$ & Sackmann  & Evolution \\
   (1992) & 3 times &   &   &   &   &   &  & et al. (\cite{Sea93}),  & of the Sun; \\
   & enhanced&    &    &   &   &   &  & Finocchi \& & chemistry of \\
   & CNO  &    &    &   &   &   &  & Gail (\cite{FG97}) & accretion discs \\
\noalign{\smallskip} \hline \noalign{\smallskip}
   %% 9th row,
   Alexander & $0 \div 1$ & + & m & + &
   + & $700 \div 10^4$ & $10^{-16} \div 10^{-5}$ &
   Hur\'{e} (\cite{Hure00}), & Accretion\\
   \& Ferguson &  &  &  &  &  &  &  &  & discs of AGN\\
   (\cite{AF94})$^{\mathrm{a; b}}$ &  &  &  &  &  &  &  &  & and YSO; \\
   &  &  &  &  &  &  &  & Driebe et & evolution \\
   &  &  &  &  &  &  &  & al. (\cite{Dea99}), & of helium \\
   &  &  &  &  &  &  &  &  & dwarfs; \\
   &  &  &  &  &  &  &  & Ikoma et & formation \\
   &  &  &  &  &  &  &  & al. (\cite{Iea01}) & of giant \\
   &  &  &  &  &  &  &  &  & planets \\
\noalign{\smallskip} \hline \noalign{\smallskip}
   %% 10th row,
   Bell \& Lin & Solar & + & a & +$^{\mathrm d}$ & -- &
   $10 \div 10^6$ & $10^{-20} \div 10^{-4}$ & Bell \& Lin &
   FU Orionis\\
   (\cite{BL94}) &  &  &  &  &  &  &  & (\cite{BL94}) & outbursts\\
\noalign{\smallskip} \hline \noalign{\smallskip}
   %% 12th row,
   Pollack et & Solar & + & -- & +$^{\mathrm c}$ & -- & $10 \div 2\,10^3$ &
   $10^{-18} \div 10^{-4}$ & Hur\'{e} (\cite{Hure00}) & Accretion \\
   al. (\cite{Pea94}) &  &  &  &  &  &  &  &  & discs\\
\noalign{\smallskip} \hline \noalign{\smallskip}
   %% 13th row,
   Seaton et & $0 \div 1$ & -- & a & + &
   + & $3\,10^3 \div 10^7$ & $10^{-15} \div 10^{-2}$ &
   Aikawa \& & Nonlinear\\
   al. (\cite{S94})$^{\mathrm{a; b}}$ &  &  &  &  &  &  &  & Antonello & pulsations\\
   (OP project) &  &  &  &  &  &  &  & (\cite{AA00a},\cite{AA00b}), &
   of Cepheids,\\
   &  &  &  &  &  &  &  & Aikawa (\cite{A01}), & accretion\\
   &  &  &  &  &  &  &  & Hur\'{e} (\cite{Hure00}) & discs\\
\noalign{\smallskip} \hline \noalign{\smallskip}
   %% 14th row,
   Henning \& & Solar & + & -- & +$^{\mathrm{c; d}}$ & --  &
   $10 \div 2\,10^3$ & $10^{-18} \div 10^{-4}$ & Bell et al. &
   Accretion\\
   Stognienko &  &  &  &  &  &  &  & (\cite{B97}) & discs\\
   (\cite{HS96})$^{\mathrm b}$ &  &  &  &  &  &  &  &  &\\
\noalign{\smallskip} \hline \noalign{\smallskip}
   %% 15th row,
   Iglesias \& & $0 \div 2\,10^{-2}$ & -- & a & + &
   -- & $5\,10^3 \div 5\,10^8$ & $10^{-15} \div 10^5$ &
   Turcotte et & Solar \\
   Rogers  &  &  &  &  &  &  &  & al. (\cite{T98}), &
   evolution,\\
   (\cite{OPAL96})$^{\mathrm{a; b}}$ &  &  &  &  &  &  &  & & \\
   (OPAL &  &  &  &  &  &  &  & Collins et &
   accretion\\
   project) &  &  &  &  &  &  &  & al. (\cite{C98}) & discs\\
\noalign{\smallskip} \hline \noalign{\smallskip}
   %% 16th row,
   Helling et & Solar, & -- & m & +$^{\mathrm c}$ & +$^{\mathrm c}$ &
   $500 \div 10^4$ & $10^{-19} \div 10^{-9}$ & Helling et &
   Dusty shells of\\
   al.~(\cite{H00})$^{\mathrm{b}}$ & LMC, &  &  &  &  &  &  &
   al.~(\cite{H00}) & long-period\\
   & SMC$^{\mathrm e}$ &  &  &  &  &  &  &  & variables \\
\noalign{\smallskip} \hline \noalign{\smallskip} 
   \end{tabular}
   \begin{list}{}{}
        \item[$^*$] atomic opacities, 
        \item[$^{**}$] molecular and atomic opacities, 
        \item[$^{\mathrm a}$] a trapezoidal table in $\rho$--$T$ parameter space,
        \item[$^{\mathrm b}$] it can be retrieved via E-mail or the Internet,
        \item[$^{\mathrm c}$] only plots are available,
        \item[$^{\mathrm d}$] analytical expressions are used,
        \item[$^{\mathrm e}$] LMC, SMC mean the element abundances of
                              the Large and the Small Magellanic Clouds
%        \item[$^{\mathrm f}$] monochromatic opacity plots are available only,
 
   \end{list}
 \end{table*}
%======================================================================
% The bibliography:
%======================================================================

\end{document}